\documentstyle[12pt]{article}

\renewcommand{\baselinestretch}{1.75} 
\newcommand{\al}{\alpha}

\newcommand{\del}{\delta}
\newcommand{\ep}{\epsilon}

\newcommand{\th}{\theta}

\newcommand{\la}{\lambda}

\newcommand{\om}{\omega}

\newcommand{\Ga}{\Gamma}
\newcommand{\De}{\Delta}

\newcommand{\La}{\Lambda}

\newcommand{\be}{\begin{eqnarray}}
\newcommand{\ee}{\end{eqnarray}}

\newcommand{\lra}{\longrightarrow}

\newcommand{\pr}{\partial}

\newcommand{\IFF}{\Longleftrightarrow}

\newcommand{\np}{\newpage}
\newcommand{\hs}{\hspace}
\newcommand{\vs}{\vspace}
\newcommand{\nl}{\newline}
\newcommand{\nn}{\nonumber}
\newcommand{\lef}{\left}

\newcommand{\bLa}{\bar{\La}}

\newcommand{\xmk}{x_k^-}

\begin{document}

\thispagestyle{empty}
\batchmode

\vs*{-25mm}
\begin{flushright}
BRX-TH-433\\[-.2in]
BOW-PH-111\\[-.2in]
HUTP-98/A39 \\
%\vs{8mm}
\end{flushright}

\begin{center}
{\Large{\bf One Instanton Predictions of a Seiberg-Witten curve from M-theory: 
the Symmetric Representation of SU($N$)}} \\
\vspace{.2in}

\renewcommand{\baselinestretch}{1}
\small
\normalsize
Isabel P. Ennes\footnote{Research supported 
by the DOE under grant DE--FG02--92ER40706.}\\
Martin Fisher School of Physics\\
Brandeis University, Waltham, MA 02254

\vspace{.1in}

Stephen G. Naculich\\
Department of Physics\\
Bowdoin College, Brunswick, ME 04011

\vspace{.1in}

Henric Rhedin\footnote{Supported by the Swedish Natural
Science Research Council (NFR),  grant no.
F--PD1--883--305.}\\
Martin Fisher School of Physics\\
Brandeis University, Waltham, MA 02254

\vspace{.1in}

Howard J. Schnitzer\footnote{Research supported in part
by the DOE under grant DE--FG02--92ER40706.}\\
Lyman Laboratory of Physics\\
Harvard University, Cambridge, MA 02138\\
and\\
Martin Fisher School of Physics\footnote{Permanent address.\\
{\tt \phantom{aaa} naculich@bowdoin.edu; 
ennes,rhedin,schnitzer@binah.cc.brandeis.edu}}\\
Brandeis University, Waltham, MA 02254

\vspace{.2in}

{\bf{Abstract}} \end{center}
\renewcommand{\baselinestretch}{1.75}
\small
\normalsize
\begin{quotation}
\baselineskip14pt
\noindent We consider N=2 supersymmetric Yang-Mills theories in four 
dimensions with gauge group SU($N$) for $N$ larger than two. Using 
the cubic curve for a matter hypermultiplet transforming 
in the symmetric representation,  obtained from M-theory 
by Landsteiner and Lopez, we calculate the prepotential 
up to the one instanton correction. We treat the  curve to be approximately 
hyperelliptic and perform a perturbation expansion 
for the Seiberg-Witten differential to get the one instanton contribution. 
We find that it reproduces the correct result for one-loop, 
and we obtain the  prediction for that curve for
the one instanton correction term.  
\end{quotation}

\np 

\setcounter{page}{1}

\section{Introduction}

Over the last few years enormous progress has been made in 
four-dimensional N=2 supersymmetric gauge  
theory. The seminal work was that of Seiberg and Witten (SW) for SU(2) pure 
Yang-Mills \cite{SeibergWitten1}, and also with matter transforming in 
the defining 
representation \cite{SeibergWitten2}. 
The main idea is 
that all the relevant information about the 
low energy limit of the theory can be obtained from an algebraic variety 
(in most cases an algebraic curve), which is taken to be a fibre over the 
moduli space of vacua. 
In particular, the low energy effective action is completely determined 
{from} a holomorphic 
function, the prepotential. This prepotential 
can be separated into three parts, a classical piece, 
a one-loop piece, and a sum of instanton contributions, and is 
calculable from the curve. 

This program was generalized by a large number of people to the classical 
groups without matter or with matter 
transforming in the defining representation \cite{Everybody}. In 
general, 
these papers do not make direct 
predictions for the prepotential, but rather they 
present a curve which may be used to obtain the prepotential, 
with suitable methods, if available.  
The conjectured curves were checked for consistency using double scaling 
limits, monodromy 
properties, decoupling limits, etc. 

The curves for the classical groups, with the above mentioned matter 
content, can all be represented 
by a double covering of the Riemann sphere, i.e. they are of hyperelliptic 
type. 
(The exceptional groups have curves of non-hyperelliptic type 
\cite{exceptional}; see also 
\cite{MartinecWarner} 
for an interesting 
account for SU($N$) pure Yang-Mills with non-hyperelliptic curves.)
A general approach to calculate instanton corrections 
to the prepotential for hyperelliptic 
curves was developed by D'Hoker, Krichever and Phong (DKP)
\cite{DHokerKricheverPhong1,DHokerKricheverPhong2}. 
These authors give explicit forms for the one and two 
instanton corrections 
for classical groups with matter in the defining representation. 

More recently, the curves of SW theory have been rederived from M-theory 
\cite{Mtheory} and also using 
geometric engineering \cite{geomenginering}. 
Working in the context of M-theory, 
Landsteiner and Lopez \cite{LandsteinerLopez} suggested curves for SW theory 
for SU($N$) 
with matter transforming in the 
symmetric as well as in the antisymmetric representations. No predictions 
for the 
prepotential were presented which could be compared with 
results obtained  from 
N=2 field theory, and hence a thorough test is desirable. 
Furthermore, the suggested curves are of non-hyperelliptic type. Although 
several consistency checks were made by Landsteiner-Lopez, there was no 
indication given as how the prepotential might be computed. 

It is the purpose of this paper to derive the prepotential from the curve 
given in \cite{LandsteinerLopez} for matter 
transforming in the symmetric representation up to the one instanton 
contribution. 
We only consider $N \geq 3$, since for 
SU(2) the symmetric representation coincides with the adjoint which is a 
scale invariant theory with four 
supersymmetries. This example is not within the scope of this paper. 

Our result should be compared with microscopic instanton 
calculations, when they become available. The importance of 
those tests is obvious. The SW method, however elegant and powerful, 
must still be tested on a case by case basis. One must of course check if 
the model can make predictions for the 
prepotential, and if it reproduces the same results as a 
calculation entirely in the context of field theory. 
By giving a prediction for the first instanton correction to the prepotential 
we provide a target for such calculations.  

The basic idea is to treat the Landsteiner-Lopez curve, which is of cubic 
form \nl $y^3+fy^2+g\bLa^2 y+\ep =0$, to be 
approximately hyperelliptic, i.e. $y^2+fy+g\bLa^2=0$. This will be a good 
approximation in the part of 
moduli space where $\ep=\bLa^6x^6$ is small, which corresponds 
to the weak coupling region. 
We then perform 
a perturbation expansion in $\ep$ for the SW differential 
to sufficient order to obtain the one instanton result. As we will see,
first order corrections will be sufficient. To determine the prepotential 
we follow the work by DKP \cite{DHokerKricheverPhong1} closely. 
Our result for the one 
instanton correction 
will, in fact, depend on residue functions $S_k(a_k)$ in the same way as 
for SU($N$) with matter in the defining representation. 
The difference will be in the detailed form of the residue functions. 

Our calculation is subjected to several consistency checks. We may 
take the mass of the hypermultiplet 
to infinity. In this limit, one should recover the result for pure Yang-Mills, 
and we show that this 
is indeed the case. 
There are also tests of the formal mathematical correctness of the result. 
First of all, for 
technical reasons, terms proportional to $\bLa$ will appear at various 
stages of 
the calculation, however, we find that they cancel in the final results, 
as they must. 
Also the dual periods, according to 
the SW ansatz, must be integrable. Since we start 
with a Riemann surface the integrability is guaranteed, and this provides 
a non-trivial test of the final 
result for the dual periods as well as for the method itself. 

One interesting fact that appears in our calculation is that the one-loop 
part of the prepotential 
follows from the hyperelliptic part of the curve by itself. Obtaining the 
one-loop part 
of the prepotential is thus not 
conclusive evidence for a particular ansatz. In our calculation, the  
evidence in favor of the 
M-theory curve (as opposed to a possible hyperelliptic candidate similar 
to the truncated M-theory curve) 
comes from the one instanton correction term to the prepotential.

\section{General Background}

We consider N=2 supersymmetric Yang-Mills theory in four dimensions with 
gauge group SU($N$), 
for $N$ larger than two, 
and with matter multiplets transforming in the symmetric rank two tensor 
representation.  

The model contains a vector multiplet corresponding to a pure gauge multiplet 
transforming in the adjoint 
representation, and matter hypermultiplets transforming in the symmetric 
representation. 
Asymptotic freedom is ensured if the beta function is  negative. 
The beta function is proportional to the 
difference of the Dynkin 
index of the adjoint representation and the sum of the Dynkin indices of 
the matter 
multiplet representations. The adjoint representation 
has index $2N$, while the symmetric representation has index $N+2$, and 
hence one flavor is the only 
asymptotically free case. 

{From} general arguments, the low 
energy effective Lagrangian is determined completely by a holomorphic 
function, the prepotential ${\cal F}$, and 
is given by
\be
{\cal L}=\frac{1}{4\pi}{\rm Im}\left(\int {\rm d}^4\th
\frac{\pr {\cal F}(A)}{\pr A_i}\bar{A_i}+
\frac{1}{2}\int {\rm d}^2\th\frac{\pr^2 {\cal F}(A)}{\pr A_i\,\pr A_j}
W^{\al}_i\,W_{\al,j}\right).
\ee 
It is known that  
the prepotential for N=2 supersymmetric theories consists of three parts; 
the classical contribution, the one-loop contribution (which corresponds 
to the only perturbative corrections by virtue of 
non-renormalization theorems), 
and non-perturbative 
instanton corrections \cite{Seiberg}. Furthermore, the prepotential must 
be of the form 
\be
& &{\cal F}=\frac{1}{2}\tau_0 A^2+\frac{i}{4 \pi}\sum_{\al \in \De_+}
(A\cdot \al)^2\,{\rm log}
\left(\frac{A\cdot \al}{\La}\right)^2\nn \\
& &-\frac{i}{8 \pi}\sum_{w \in W_G}\sum_{j=1}^{N_f}(A\cdot w+m_j)^2\,
{\rm log}\left(\frac{A\cdot w+m_j}{\La}\right)^2+
\sum_{d=1}^{\infty}{\cal F}_d(A)\La^{(2N-I_D)d}. \label{prepotential}
\ee
In this expression,  $\De_+$ denotes the set of positive roots, $W_G$ is the 
set of weights, ${\cal F}_d$ are 
instanton corrections, $I_D$ denotes the sum of the Dynkin indices of the 
matter hypermultiplet representations, 
$\La$ is the dynamically generated scale of the theory, $N_f$ is the number 
of matter hypermultiplets, and $m_j$ are the respective masses. Furthermore, 
$A_i$ is an N=1 chiral superfield whose scalar 
components $e_i$ parametrize the flat directions of the potential 
\be
V(\phi)=Tr[\phi,\phi^{\dagger}]^2.
\ee
Here $\phi$ is the scalar field of the N=2 vector supermultiplet 
transforming in the adjoint representation. \\

The SW ansatz postulates that the prepotential ${\cal F}(a)$ 
can be obtained from a fibration of algebraic varieties over the 
moduli space of vacua, where 
$a_i$ is the quantum corrected $e_i$. 
Associated with each fiber 
there is a preferred differential $\la$, the SW differential. Using a 
canonical set of 
one cycles $(A_i,B_i)$ that 
form a basis for the first homology group, 
the periods $a_k$ and their dual partners $a_{D,k}$ are obtained from the 
curve 
and $\la$ by means of the period integrals  
\be
2\pi i a_k=\oint_{A_k}\la \hs{15mm} 2\pi ia_{D,k}=\oint_{B_k}\la. 
\ee
The prepotential is related to the periods by  
\be
a_{D,k}=\frac{\pr {\cal F}}{\pr a_k}.\label{patata}
\ee

To verify the validity of the ansatz, one must perform microscopic instanton 
calculations similar to those in ref. \cite{instanton}. 
Those calculations will provide the correct (i.e. 
obtained from the underlying field theory) 
instanton 
coefficients ${\cal F}_d(a)$. The purpose of this paper 
is to derive the one instanton contributions to the prepotential, from 
the curve deduced by 
Landsteiner and Lopez from M-theory for the symmetric representation of 
$SU(N)$. 
This will provide a prediction that can be checked against microscopic 
instanton 
calculations, when they become available.

\section{The Curve}

The curve proposed from M-theory considerations by Landsteiner and Lopez 
for matter transforming in the symmetric representation 
takes the form \cite{LandsteinerLopez}
\be
y^3+f(x)y^2+g(x)\bLa^2 y+\ep (x)=0, \label{curve}
\ee
where
\be
f(x)=\prod_{i=1}^N (x-e_i) \hs{15mm} g(x)=(-1)^N x^2\prod_{i=1}^N (x+e_i) 
\hs{15mm} \ep(x)=\bLa^6 x^6. 
\label{deffge}
\ee
Here $\bLa^2=\La^{N-2}$, and $e_i$ are the classical moduli which 
parametrize the weights of the defining representation of SU($N$). 
Asymptotic freedom requires the number of 
flavors to be at most one, and the classical order parameters satisfy 
\be
\sum_{i=1}^N e_i=0, \label{mass}
\ee
for the massless hypermultiplet. 

It can be checked that the curve (\ref{curve}) is invariant under the 
involution
\be
y\lra\frac{\bLa^4x^4}{y} \hs{10mm} x\lra -x. \label{involution}
\ee
The existence of an involution in this case reflects the fact that the 
genus of 
the algebraic variety corresponding to the 
curve (\ref{curve}) is larger than the dimension of the moduli space.

\begin{picture}(500,275)(10,10)

%horizontal sheets 

\put(55,275){\line(1,0){380}}
\put(5,225){\line(1,0){380}}
\put(5,225){\line(1,1){50}}
\put(385,225){\line(1,1){50}}

\put(55,175){\line(1,0){380}}
\put(5,125){\line(1,0){380}}
\put(5,125){\line(1,1){50}}
\put(385,125){\line(1,1){50}}

\put(55,75){\line(1,0){380}}
\put(5,25){\line(1,0){380}}
\put(5,25){\line(1,1){50}}
\put(385,25){\line(1,1){50}}

%markings

\put(417,247){$y_1$}
\put(417,147){$y_2$}
\put(417,47){$y_3$} 

\put(212,37){$0$}

\put(80,37){$-e_{N-1}$}
\put(153,37){$-e_2$}
\put(180,37){$-e_1$}
\put(367,37){$-e_N$}

\put(55,255){$e_N$}
\put(237,255){$e_1$}
\put(262,255){$e_2$}
\put(337,255){$e_{N-1}$}

%vertical lines

\put(340,150){\line(0,1){100}}
\put(265,150){\line(0,1){100}}
\put(240,150){\line(0,1){100}}
\put(60,150){\line(0,1){100}}

\put(215,50){\line(0,1){2}}
\put(215,58){\line(0,1){5}}
\put(215,68){\line(0,1){5}}
\put(215,78){\line(0,1){5}}
\put(215,88){\line(0,1){5}}
\put(215,98){\line(0,1){5}}
\put(215,108){\line(0,1){5}}
\put(215,118){\line(0,1){5}}
\put(215,128){\line(0,1){5}}
\put(215,138){\line(0,1){5}}
\put(215,148){\line(0,1){2}}

\put(90,50){\line(0,1){100}}
\put(165,50){\line(0,1){100}}
\put(190,50){\line(0,1){100}}
\put(375,50){\line(0,1){100}}

%dotts

\put(110,47){$\cdot$}
\put(125,47){$\cdot$}
\put(140,47){$\cdot$}

\put(315,147){$\cdot$}
\put(300,147){$\cdot$}
\put(285,147){$\cdot$}

\put(110,147){$\cdot$}
\put(125,147){$\cdot$}
\put(140,147){$\cdot$}

\put(315,247){$\cdot$}
\put(300,247){$\cdot$}
\put(285,247){$\cdot$}

\end{picture}
\begin{center}
{\bf Fig.1: The sheet structure for the cubic curve.}
\end{center}

The curve (\ref{curve}) 
can be regarded as a three-sheeted branched covering of the Riemann sphere 
parametrized by the complex 
coordinate $x$.  It describes a Riemann surface of genus $2N-2$. 
If we change the variable $y$ to  $w=y+f(x)/3$, the curve becomes  
$w^3+a(x)w+b(x)=0$. %In these new variables, the roots are
%\be
%w_1=&p_++p_- \nn \\ 
%w_2=-\frac{1}{2}(p_++p_-) + i\frac{\sqrt{3}}{2}(p_+-p_-) 
%\hs{2mm}, & & \hs{1mm}  w_3=-\frac{1}{2}(p_++p_-)-i\frac{\sqrt{3}}{2}
%(p_+-p_-) \nn \\
%p_{\pm}(x)=\left(-\frac{b(x)}{2}\pm\sqrt{s(x)}\right)^{1/3} &\ & \hs{10mm} 
%s(x)=\frac{a^3}{27}+\frac{b^2}{3}, 
%\label{roots}
%\ee
%where $s(x)$ is the discriminant of the curve. 
To find the positions of the ramification points of the curve 
we look for the points where two sheets coincide, i.e. 
when $w_i=w_j$ for $i\neq j$, which implies that the discriminant 
$s(x)=0$, where 
\be
s(x)=\frac{a^3(x)}{27}+\frac{b^2(x)}{4}=-\frac{\bLa^4}{108}\left(f^2g^2+
{\cal O}(\bLa^2)\right). \label{roots}
\ee 
To zeroth order in $\bLa$, the solutions to  the equation  $s(x)=0$ 
correspond to the 
set $\{ 0, \pm e_i\}$. 
The corrections to these solutions will be such that the ramification 
points $\pm e_i$ 
split into pairs for non-vanishing $\bLa$. 
In the perturbative regime, 
i.e. for $\bLa\ll e_i, \,\, i=1,...,N$, one has square root cuts over 
$\pm e_i\,$, while at $x=0$ 
there are no cuts at all. A careful 
analysis of how the phase changes as one moves between the 
ramification points shows that sheets two and three are connected over 
$-e_i\,$,  while sheets one and two 
are connected over $e_i\,$. This structure is depicted in fig.1. 
The dashed line over $x=0$ indicates that sheets two and three  
touch here,   but there are no branch cuts at $x=0$.  
The other vertical lines indicate the location of the square root 
cuts, which we choose to run between $x^{\pm}_i$. (Thus $x^{\pm}_i\lra e_i$ 
as $\bLa\lra 0$.) 

The structure of the Riemann surface may also be obtained in a more intuitive 
way from the involution 
of the curve. Since the curve (\ref{curve})
is invariant under the involution (\ref{involution}),  the set of roots is at 
most permuted by the involution. In fact, one finds that the involution 
exchanges sheets one and 
three, and maps sheet two onto itself. 
This is easy to verify by using the action of the 
involution on the roots, and computing to the lowest order in 
$\bLa$, where the roots are given by 
\be
y_1=-f+... \hs{10mm} y_2=-\frac{\bLa^2g}{f}+... \hs{10mm} y_3=
-\frac{\bLa^4x^6}{g}+...
\ee
Since the involution includes the discrete map $x\lra -x$, the 
sheet structure follows. \\

The strategy of this paper is to make a systematic expansion of the solutions 
of the curve around the hyperelliptic approximation, which induces a 
perturbation 
series for the SW differential. We will keep terms to sufficient accuracy  
to obtain 
the one instanton contribution to the prepotential. 
The approximation scheme means that only a two-sheeted structure will appear 
in the calculation. 
The zeroth approximation to the curve
(\ref{curve}) is the hyperelliptic equation
\be
y^2+yf(x)+\bLa^2g(x)=0 \label{hyperelliptic}, 
\ee
with roots 
\be
y_{\pm}=-\frac{f(x)}{2}\pm\left[\frac{(f(x))^2}{4}-\bLa^2 g(x)\right]^{1/2}
\equiv-\frac{f(x)}{2}\pm r(x).
\label{heroots}
\ee
There are square root branch cuts over $e_i$ connecting sheets one and 
two. When computing the 
periods and the dual periods, we only need to consider cycles related to 
those 
branchings, since the approximation only involves these two sheets.  
(Even when we include perturbations in $\ep$, only the same cycles are 
required. This means that 
this is indeed a projection onto a subvariety.)
Notice that the hyperelliptic approximation breaks down near $x=0$. However, 
this does not 
change the dual periods.

To find the correction terms to the roots of the cubic equation 
(\ref{curve}) as well as to the 
SW differential 
\be
\la=x\frac{{\rm d}y}{y}, \label{swdifferential}
\ee
we make a pertubation expansion around the last term of the cubic equation, 
that is, around 
$\ep (x)$. 
Expand solutions to (\ref{curve}) in powers of $\ep$,  
\be
y_i(x)=\sum_{n=0}^{\infty}\al^{(i)}_n \ep^n, \hs{5mm} (i=1,2,3)
\label{epseries}
\ee
and insert (\ref{epseries}) into $\prod^3_{i=1}(y-y_i)=0$. One may  
solve for the coefficients $\al_n^{(i)}$ iteratively order by order in 
the perturbation theory 
by comparing this expression with (\ref{curve}). 
We find, correct to first order in $\ep$, the roots
\footnote{See ref.
\cite{NaculichRhedinSchnitzer}, Appendix A for a more explicit account of 
the pertubation expansion.}
\be
& &y_1(x)=-\left(\frac{f}{2}+r\right)\left[1+\ep\left(\frac{1}{2\bLa^2 g r}+
\frac{r-f/2}{\bLa^4 g^2}\right)\right]+...\nn\\
& &y_2(x)=-\left(\frac{f}{2}-r\right)\left[1+\ep\left(\frac{-1}{2\bLa^2 g r}-
\frac{r+f/2}{\bLa^4 g^2}\right)\right]+... \nn \\
& &y_3(x)=-\frac{\ep}{\bLa^2g}+...  \label{rootexpansion}
\ee

The perturbation expansion enables us to express
the SW differential for sheet one to first order in $\ep (x)=\bLa^6x^6$ as 
\be
\la_1=\frac{x}{r+f/2}{\rm d}\left(r+\frac{f}{2}\right)+x{\rm d}
\left[\ep\left(\frac{1}{2 \bLa^2 gr}+
\frac{r-f/2}{\bLa^4 g^2}\right)\right]\,.
\label{swdpert}
\ee
The first term corresponds to the differential for the hyperelliptic 
approximation. 
Although this 
term has the standard form, notice that the function $g(x)$ in our case 
is moduli dependent, which is different than SW theory for SU($N$) with 
no matter or matter in the defining 
representation. 
We will find that the hyperelliptic approximation correctly predicts 
the one-loop part of the prepotential. 
However, the ${\cal O}(\ep)$ correction to the hyperelliptic 
approximation will be important for our calculation of 
the one instanton contribution to the prepotential.

For later reference, note that the weights of the symmetric representation 
are 
\be
e_i+e_j \hs{10mm} i\leq j 
\hs{10mm}i,j=1,...,N.
\ee

\section{The Periods}

The purpose of this paper is to compute the explicit form 
of the prepotential up to the one instanton contribution for the $N=2$ 
supersymmetric
Yang-Mills theory with matter transforming in the symmetric representation 
{from} the 
Landsteiner Lopez curve. In
order to do that, we must calculate the explicit expression of the periods 
$a_k$ and
the dual periods $a_{D,k}$ up to order $\bLa^2$. 

As the curve (\ref{curve}), proposed from M-theory for the symmetric 
representation, 
is not hyperelliptic, we develop a systematic procedure which generalizes 
previous methods. This involves 
a systematic expansion of the solution  to (\ref{curve}) about the 
hyperelliptic 
approximation to this curve, which then induces similar corrections to 
the SW differential $\la$.  We will show that the corrections to 
${\cal O}(\ep)$ in $\la$ do not 
contribute to the result of the periods $a_k$, although they are  
important in the computation of the dual periods $a_{D,k}$.

For the  computation of the periods, we will work on the first sheet and 
consider the SW differential 
given in eq. (\ref{swdpert}). 
We arrange the branch cuts to extend from  $x^-_k$ to $x^+_k$, where 
$x^{\pm}_k\lra e_k$ as 
$\bLa\lra 0$. We describe the cycles after projecting onto the sub-variety 
i.e. that of the 
hyperelliptic approximation. 
The $A_k$ cycles are then closed contours encircling the branch cuts with 
centers 
around $e_k$ (see fig.1). The $B_k$ 
cycles consist of curves which go from $x^-_1$ to $x^-_k$ on sheet one, 
pass through the 
branch cut between $x_k^-$ and $x_k^+$, and return to $x_1^-$ on sheet 
two. It is 
clear that the intersection number of the cycles $A_i$ and $B_j$ will be 
proportional to 
$\del_{ij}$. 

First of all, we will evaluate the portion of the periods that comes from 
the hyperelliptic approximation. 
For this purpose, define
\be
\la\,=\,\la_I\,+\la_{II}+...,
\ee
where the first term refers to the hyperelliptic approximation, and the 
second corresponds to the
first-order corrections in $\ep$.
In order to do so, express the $\ep=0$ part of the SW differential on 
sheet one, eq. (\ref{swdpert}),  as
\be
(\la_1)_{I}=\frac{x{\rm d}x}{2r}\left(f'-\frac{g'}{2g}f\right)+
\frac{x{\rm d}x}{2g}g',
 \label{seiberg}
\ee
where $f'$ and $g'$ denote $\pr f/{\pr x}$ and $\pr g/{\pr x}$, respectively. 
The last term in  (\ref{seiberg}) can be neglected  in the computation of 
the periods $a_k$,
since it does not have any residues within the contour $A_k$. Following 
DKP \cite{DHokerKricheverPhong1}, the integral  we need to 
evaluate is \np
\be
2\pi i (a_k)_I&\equiv&\oint_{A_k}{\rm d}xx\frac{\left(\frac{f'}{f}-
\frac{g'}{2g}\right)}
{\sqrt{1-4\bLa^2g/f^2}} \nn \\
=\oint_{A_k}{\rm d}xx\frac{f'}{f}&+&\sum_{m=1}^{\infty}
\frac{\Ga(m+1/2)}{\Ga(m+1)\Ga(1/2)}(4\bLa^2)^m 
\oint_{A_k}{\rm d}xx\left(\frac{f'}{f}-\frac{g'}{2g}\right)
\frac{g^m}{f^{2m}}, 
\label{integral}
\ee
where we have expanded the denominator in a convergent power series 
in $\bLa^2$. 
The first integral in (\ref{integral}) can be easily evaluated, with 
the result   
\be
\oint_{A_k}{\rm d}xx\frac{f'}{f}=
2\pi i e_k.
\ee
For the remaining part of the integral (\ref{integral}),  we use the 
following identity 
\cite{DHokerKricheverPhong1}
\be
x\left(\frac{f'}{f}-\frac{g'}{2g}\right)\frac{g^m}{f^{2m}}=
-\frac{\rm d}{{\rm d} x}\left(\frac{xg^m}{2mf^{2m}}\right)
+\frac{1}{2m}\frac{g^m}{f^{2m}}. \label{idtwo}
\ee
The total derivative in (\ref{idtwo}) drops out, and for the remainder we 
only need 
the contribution with $m=1$,  since we are only interested in 
$\bLa^2$ corrections. Introduce the residue functions $S_k(x)$ defined by 
\be
\frac{4g}{f^2}\equiv\frac{S_k(x)}{(x-e_k)^2},
\ee
where
\be
S_k(x)=\frac{4(-1)^Nx^2\prod_{i=1}^N(x+e_i)}
{\prod_{i\neq k}(x-e_i)^2}. \label{residuefunc}
\ee
Using the definition of $S_k(x)$ and that 
${\Ga(3/2)\over{\Ga(2)\Ga(1/2)}}\,=\,1/2, $ 
we can conclude that the final expression for the 
periods $a_k$, up to one-instanton 
order is
\be
a_k=e_k+\frac{\bLa^2}{4}\frac{\pr S_k}{\pr x}(e_k). \label{periods}
\ee

Equation (\ref{periods}) is not changed 
to ${\cal O}(\bLa^2)$ by the ${\cal O}(\ep)$ corrections to the SW 
differential, as can 
be checked  directly from (\ref{swdpert}). Indeed, the first term 
$\ep/(2\bLa^2r)$ is of 
higher order, 
while the remaining terms do not have residues within the $A_k$ contours. 
In order to see that 
$\ep/(2\bLa^2r)$ is 
higher order than $\bLa^2$, we notice that its lowest order 
contribution is 
$2\bLa^4x^6/f(x)$. Although we 
 can expand this factor by 
 using partial fractions,  
we will obviously never get contributions to the one instanton term.

\section{The Dual Periods}

One evaluates the SW differential for the $B_k$ cycles by means of a 
contour that goes 
{from} $x_1^-$ to $x^-_k$ on  
sheet one, crosses the branch cut at $e_k$ to sheet two, runs back 
{from} $x_k^-$ to $x_1^-$ on sheet two, and passes to sheet one through 
the branch cut 
at $e_1$. The SW differential on sheet two is obtained by 
taking $r\lra -r$ in $\la_1$. The $B_k$ cycles require the difference 
$\la_1-\la_2$ in the integral   
therefore only terms with odd powers of  $r$ 
will contribute to the dual periods.  

The dual periods $a_{D,k}$ have two different contributions; the first 
corresponds 
to the hyperelliptic approximation, while the second comes from the 
correction to the hyper- elliptic approximation.
They can be expressed as
\be
a_{D,k}\,=\,(a_{D,k})_I\,+\,(a_{D,k})_{II}.
\ee
In order to find the contribution coming from the hyperelliptic 
approximation, 
we need to introduce a 
regularization parameter 
$\xi$, which allows us to expand $r^{-1}$ in the integral for the dual 
periods, as was argued 
by DKP \cite{DHokerKricheverPhong1}. In what follows,  it is 
understood that the limit $\xi\lra 1$ is taken in the final results. Then 
the hyperelliptic approximation for the dual periods is given 
by
\be
2\pi i (a_{D,k})_I=2\sum_{m=0}^{\infty}\frac{\Ga(m+1/2)}{\Ga(m+1)\Ga(1/2)}
\xi^{2m}(4\bLa^2)^m 
\int^{x^-_k}_{x_1^-}{\rm d}xx\left(\frac{f'}{f}-\frac{g'}{2g}\right)
\frac{g^m}{f^{2m}}. \label{dualstart} 
\ee 
For notational convenience 
we will suppress the dependence on $x_1^-$, which we will restore at 
the end. 

In order to truncate the series at the desired order in $\bLa$, one must 
know $x^-_k$ to 
sufficiently high order. By definition, $x^-_k$ satisfies $s(x^-_k)=0$. 
The explicit form of 
the discriminant, defined in 
eq. (\ref{roots}), in terms of the functions $f$ and $g$, is 
\be
s(x)=-\frac{\bLa^4}{108}\left[f^2g^2-4\bLa^2(f^3x^6+g^3)+
18\bLa^4x^6fg-27\bLa^8x^{12}\right].
\ee
To order $\bLa^2$, this reduces to the identity 
$f^2(x_k^-)-4\bLa^2g(x_k^-)=0$, which is 
appropriate to the hyperelliptic approximation, i.e. $r(x_k^-)=0$ in 
(\ref{heroots}). This is derived in an alternate way
in ref. \cite{NaculichRhedinSchnitzer}.
After a Taylor expansion, we get to one instanton accuracy
\be
\xmk=e_k-\bLa S^{1/2}_k(e_k)+\frac{\bLa^2}{2}
\frac{\pr S_k}{\pr x}(e_k)+... \label{branchpoints}
\ee

Since some integrations in eq. (\ref{dualstart}) produce inverse powers 
of $\bLa$, 
we need to examine all orders of $m$ in eq. (\ref{dualstart}).  
For future reference, introduce the sum 
\be
H(\om,n)\equiv \sum_{m=n}^{\infty}\frac{\Ga(m+1/2)}{\Ga(m+1)\Ga(1/2)}\om(m).
\ee
The relevant functions $\om(m)$ and values of $H(\om,n)$ for our 
calculation are presented in Table ~1, which we have extracted from DKP 
\cite{DHokerKricheverPhong1}. 
\begin{center}
\begin{tabular}{|c|c|c|}    
\hline
$\om(m)$ & $n$ & $H$ \\
\hline
$1/(2m)$ & 1 & {\rm log}2 \\
$1/[2(m-1)]$ & 2 & (1/2){\rm log}2+1/4 \\
$1/[2m(2m-1)]$ & 1 & -{\rm log}2+1 \\
\hline 
\end{tabular} \\
\ \\
\bf{Table 1}
\end{center}

First, consider  the $m=0$ contribution to the dual period (\ref{dualstart})
\be
2\int^{\xmk}{\rm d}xx\left(\frac{f'}{f}-\frac{g'}{2g}\right)=
2\int^{\xmk}{\rm d}xx\left[\sum_{i=1}^N\,\frac{1}{x-e_i}-
\frac{1}{x}-\sum_{i=1}^N\,\frac{1}{2(x+e_i)}\right]\nn \\
=(N-2)\xmk+2\sum_{i=1}^Ne_i\,{\rm log}(\xmk-e_i)+\sum_{i=1}^Ne_i\,
{\rm log}(\xmk+e_i)  \label{zerocontrib}.
\ee

By using  eq. (\ref{idtwo}), the  integral (\ref{dualstart}) for all 
$m\geq 1$ 
can be expressed in a more 
convenient form. 
The total derivative can be evaluated by recalling that 
$4g(\xmk)/f^2(\xmk)\!~=~\!1$ 
to our order of accuracy, and the  
prefactors of this term sum to 
$H(1/m,1)=2\,{\rm log}2$, according to Table 1. 
Hence, the  contribution coming from the $m \neq 0$ terms  in eq. 
(\ref{zerocontrib}) becomes
\be
& &2\sum_{m=1}^{\infty}\frac{\Ga(m+1/2)}{\Ga(m+1)\Ga(1/2)}\xi^{2m}(4\bLa^2)^m 
\int^{x^-_k}{\rm d}xx\left(\frac{f'}{f}-\frac{g'}{2g}\right)
\frac{g^m}{f^{2m}}\nn\\
& &=-2\xmk\,{\rm log}2+
2\sum_{m=1}^{\infty}\frac{\Ga(m+1/2)}{\Ga(m+1)\Ga(1/2)}
\frac{\xi^{2m}(4\bLa^2)^m}{2m}\int^{\xmk}{\rm d}x\frac{g^m}{f^{2m}}.
\label{totdercontrib}
\ee

To evaluate the remaining integral we need to be careful to keep all 
terms contributing to 
order $\bLa^2$, as the integration may produce inverse powers 
of $\bLa$. 
Because of this, we will have to deal with an infinite power series in 
$m$. In 
order to carry out these integrations, we use partial fractions for the 
integrand, i.e. 
\be
\frac{4^mg^m}{f^{2m}}\equiv\sum_{i=1}^N\sum_{p=1}^{2m}
\frac{Q^{(2m)}_{i,p}}{(x-e_i)^p},
\label{thirtyone}
\ee
for some suitable constants $Q^{(2m)}_{i,p}$. To study the relevant 
contributions of eq.
(\ref{thirtyone}) to (\ref{totdercontrib}) we need consider three 
separate cases. 
The first possibility appears when $i=k$. In this case,
since $(\xmk-e_k)$  is of order $\bLa$, only $p=2m$ and $p=2m-1$ will 
contribute to order $\bLa^2$.   
However, all $m$ will contribute to the sum in (\ref{totdercontrib}) 
for these values of $p$. 
For example, for $p=2m$, we obtain $\bLa^{2m}/(\xmk-e_k)^{2m-1}$ 
after performing the integration, which 
provides contributions of order $\bLa$ for all $m\geq 1$. 

The second possibility to consider is $i\neq k$. In this case, the factor 
$\xmk-e_i$ is of zeroth order
in $\bLa$, so $m=1$ will be the only relevant contribution to order 
$\bLa^2$ in 
(\ref{totdercontrib}).

Finally the value $p=1$ 
requires special attention in our discussion, since the contribution 
(\ref{totdercontrib}) in this case will generate logarithmic terms,
both for $i=k$ and for $i\neq k$.

First, let us turn our attention to $p=1$. Notice from (\ref{thirtyone}) 
that
$Q^{(2m)}_{i,1}$ is nothing but the residue of the function 
$(4g)^m/f^{2m}$. 
Hence 
\be
Q^{(2m)}_{i,1}=\frac{2m}{2\pi i}\oint_{A_k}{\rm d}x x\left(\frac{f'}{f}-
\frac{g'}{2g}\right)\frac{4^mg^m}{f^{2m}}.
\ee
Also 
\be
& &\frac{1}{2\pi i}\sum_{m=1}^{\infty}
\frac{4^m\bLa^{2m}\Ga(m+1/2)}{\Ga(m+1)\Ga(1/2)}
\oint_{A_k}{\rm d}xx\left(\frac{f'}{f}-
\frac{g'}{2g}\right)\frac{g^m}{f^{2m}} \nn \\
& &=\frac{1}{2\pi i}\oint_{A_k}{\rm d}xx\left(\frac{f'}{f}-
\frac{g'}{2g}\right)\left(\frac{1}{\sqrt{1-4\bLa^2g/f^2}}-1\right) \nn \\
& &=a_k-e_k.
\ee
Taking into account that the integrand for $p=1$ is $1/(x-e_i)$,  
the contribution 
of this  term to (\ref{totdercontrib}) is
\be
& &2\sum_{m=1}^{\infty}\sum_{i=1}^N\frac{\Ga(m+1/2)}{\Ga(m+1)\Ga(1/2)}
\frac{\xi^{2m}\bLa^{2m}}{2m}
\int^{x^-_k}{\rm d}x\frac{Q^{(2m)}_{i,1}}{x-e_i}\nn \\
& &=2\sum_{i=1}^{N}(a_i-e_i)\,{\rm log}(\xmk-e_i). \label{ponecontrib}
\ee

As we have argued, for $i\neq k$, only $m=1$ will be relevant to order 
$\bLa^2$. 
Its contribution to (\ref{totdercontrib}) 
becomes
\be
& &2\sum_{i\neq k}\sum_{m=1}^{\infty}
\sum_{p=2}^{2m}\frac{\Ga(m+1/2)}{\Ga(m+1)\Ga(1/2)}
\frac{\xi^{2m}\bLa^{2m}}{2m} 
\int^{x^-_k}{\rm d}x\frac{Q^{(2m)}_{i,p}}{(x-e_i)^p}\nn \\
& &=-\frac{1}{2}\,\bLa^2\sum_{i\neq k}\frac{Q^{(2)}_{i,2}}{x^-_k-e_i}+
{\cal O}(\bLa^3).  \label{icontrib}
\ee

For   $m \neq 0$, our integral thus becomes
\be
& &2\sum_{m=1}^{\infty}
\frac{\Ga(m+1/2)}{\Ga(m+1)\Ga(1/2)}\,\frac{\xi^{2m}}{2m}(4\bLa^2)^m 
\int^{\xmk}{\rm d}x\frac{g^m}{f^{2m}} \\
& &=2\sum_{i=1}^{N}(a_i-e_i){\rm log}(\xmk-e_i)
-\frac{1}{2}\,\bLa^2\sum_{i\neq k}\frac{Q^{(2)}_{i,2}}{x^-_k-e_i}\nn \\
& &-2\sum_{m=1}^{\infty}\frac{\Ga(m+1/2)}{\Ga(m+1)\Ga(1/2)}
\frac{(\bLa\xi)^{2m}}{2m}\left[
\,\frac{1}{(2m-1)}\,\frac{Q^{(2m)}_{k,2m}}{(\xmk-e_k)^{2m-1}}+
\frac{\th_{m-2}}{(2m-2)}\,\frac{Q^{(2m)}_{k,2m-1}}{(\xmk-e_k)^{2m-2}}\right].
%+{\cal O}(\bLa^3) 
\nn \label{Howard}
%\nn \\
%\!\!\!\!\!\! & &-2\sum_{m=2}^{\infty}\frac{\Ga(m+1/2)}{\Ga(m+1)\Ga(1/2)}\,
%\frac{\xi^{2m}}{2m}(4\bLa^2)^m
%\,\frac{1}{(2m-2)}\,\frac{Q^{(2m)}_{k,2m-1}}{(\xmk-e_k)^{2m-2}}+
%{\cal O}(\bLa^3). \label{Howard}
\ee
to one instanton order, and where $\th_s=1$ for $s\geq 0$, and $\th_s=0$ 
for $s<0$. 

In order to obtain the final expression for the integral,  we must 
know the partial fraction coefficients $Q^{(2m)}_{i,p}$. These terms
can be obtained  from the  evaluation of $(4g)^m/ f^{2m}$ 
near $x=e_i$. We find
\be
\frac{4^mg^m}{f^{2m}}=\left(\frac{S_k(x)}{(x-e_k)^2}\right)^m
\lra
\frac{S_k^m(e_k)}{(x-e_k)^{2m}}+
\frac{mS_k^{m-1}(e_k)\frac{\pr S_k}{\pr x}(e_k)}{(x-e_k)^{2m-1}}+...
\ee
which, by identification,  gives $Q^{(2m)}_{k,2m}=S^m_k$ and 
$Q^{(2m)}_{k,2m-1}=mS^{m-1}_k\frac{\pr S_k}{\pr x}(e_k)$. Using 
this result, Table 1 for the $m$ summation, 
and a Taylor expansion of $1/(\xmk-e_k)$ with the explicit form of 
$\xmk$ c.f. 
eq. (\ref{branchpoints}), the total contribution for $p=2m$ and $p=2m+1$ 
{from} eq. (\ref{Howard}) 
when $i=k$ becomes
\be
& &-2\sum_{m=1}^{\infty}\frac{\Ga(m+1/2)}{\Ga(m+1)\Ga(1/2)}\,
\frac{(\bLa\xi)^{2m}}{2m}
\left[\frac{1}{(2m-1)}\,\frac{Q^{(2m)}_{k,2m}}{(\xmk-e_k)^{2m-1}}+
\frac{\th_{m-2}}{(2m-2)}\,\frac{Q^{(2m)}_{k,2m-1}}{(\xmk-e_k)^{2m-2}}\right]
\nn\\
%\!\!\!\!\!\! & &-2\sum_{m=2}^{\infty}\frac{\Ga(m+1/2)}{\Ga(m+1)\Ga(1/2)}\,
%\frac{\xi^{2m}}{2m}(4\bLa^2)^m
%\,\frac{1}{(2m-2)}\,\frac{Q^{(2m)}_{k,2m-1}}{(\xmk-e_k)^{2m-2}} \nn\\
& &=2\bLa(1-{\rm log}2)S_k^{1/2}(e_k)+\bLa^2(\frac{1}{2}{\rm log}2-
\frac{1}{4})
\frac{\pr S_k}{\pr x}(e_k)+ {\cal O}(\bLa^3)
.\label{kcontrib}
\ee

Equation (\ref{zerocontrib}) and the relevant parts of 
(\ref{totdercontrib}), (\ref{ponecontrib}), 
(\ref{icontrib}), and (\ref{kcontrib}) add to 
\be
2\pi i(a_{D,k})_I=(N-2-2{\rm log}2)\xmk+2\sum_{i=1}^Na_i\,
{\rm log}(\xmk-e_i)+
\sum_{i=1}^Ne_i\,{\rm log}(\xmk+e_i) \nn \\
+2\bLa(1-{\rm log}2)S_k^{1/2}(e_k)+\bLa^2\left[-\frac{1}{2}
\sum_{i\neq k}\frac{S_i(e_i)}{\xmk-e_i}+
(\frac{1}{2}{\rm log}2-\frac{1}{4})\,\frac{\pr S_k}{\pr x}(e_k)\right]+
{\cal O}(\bLa^3).\label{rrr}
\ee

This is not yet the expression for the dual periods in 
the hyperelliptic approximation. 
We must express $x^-_k$ and $e_i$ in terms of $a_k$ and $a_i$ respectively. 
Also notice that we have contributions proportional to 
$\bLa$ in (\ref{rrr}). This is, of course, unphysical since the one 
instanton contribution 
corresponds to terms of order $\bLa^2$. 
We have verified that 
the next term in perturbation theory in $\ep(x)$ will contribute only to 
one instanton order and higher, hence the terms of order $\bLa$ 
must cancel in the hyperelliptic approximation. 

The calculation, being tedious and relatively straightforward, can, 
however, be 
facilitated by a few useful rearrangements, which we quote in 
what follows. We then state the result for the dual periods obtained from 
the zeroth order approximation in $\ep$ in the perturbation theory.  
Additional intermediate results are given in Appendix A. 

First of all,  notice that from the expression of the prepotential, 
we expect the dual periods at order $(\bLa)^0$ to 
contain terms of the type $(a_k-a_i)\,{\rm log}(a_k-a_i)$ as well as \nl 
$(a_k+a_i)\,{\rm log}(a_k+a_i)$. With this in mind, we use the following 
identity that is valid to order $\bLa^2$
\be
f(\xmk)=-2\bLa g^{1/2}(\xmk) \hs{3mm} \IFF \hs{3mm} 
\prod^N_{i=1}(\xmk-e_i)=-2\bLa\xmk(-1)^{N\over 2}
\prod^N_{i=1}(\xmk+e_i)^{1/2}.
\ee
Taking the logarithm yields 
\be
0={\rm log}(-\bLa)+{\rm log}2+\frac{1}{2}{\rm log}(-1)^N-
\sum^n_{i=1}{\rm log}(\xmk-e_i)+\frac{1}{2}\sum^n_{i=1}{\rm log}(\xmk+e_i)
+{\rm log}\xmk.
\label{zero} 
\ee
Multiplying (\ref{zero}) by $2\xmk$ and adding it to 
$2\pi i(a_{D,k})_I$, we obtain \np
\be
& &2\pi i(a_{D,k})_I\nn \\
& &=[N-2+2\,{\rm log}(-\bLa)+{\rm log}(-1)^N]\xmk-2\sum_{i=1}^N(\xmk-a_i)
{\rm log}(\xmk-e_i)\nn \\ 
& &+\sum_{i=1}^N(\xmk+e_i)\,{\rm log}(\xmk+e_i)+
2\xmk\,{\rm log}\xmk+
2\bLa(1-{\rm log}2)S_k^{1/2}(e_k)\nn \\ 
& &+\bLa^2\left[-\frac{1}{2}\sum_{i\neq k}\frac{S_i(e_i)}{\xmk-e_i}+
(\frac{1}{2}{\rm log}2-\frac{1}{4})\frac{\pr S_k}{\pr x}(e_k)\right]. 
\label{finalint}
\ee

To evaluate the terms of the type $(\xmk-a_i)\,{\rm log}(\xmk-e_i)$ 
for $i\neq k$, we use the 
following expansion
\be
{\rm log}(\xmk-e_i)\approx 
{\rm log}(e_k-e_i)+\frac{\xmk-e_i}{e_k-e_i}-\frac{1}{2}
\frac{(\xmk-e_i)^2}{(e_k-e_i)^2}.
\label{idthree} 
\ee

Taking into account eqs. (\ref{zero}), (\ref{finalint}), and 
the results in Appendix A,  we obtain  the classical and one-loop  
contribution to the dual periods in the form
\be
& &2\pi i\left[(a_{k,D})_{cl}+(a_{k,D})_{1-loop}\right]_I=[N-2+2\,
{\rm log}(-\bLa)+{\rm log}(-1)^N]a_k\nn \\ 
& &-2\sum_{i\neq k}(a_k-a_i)\,
{\rm log}(a_k-a_i)+
\sum_{i=1}^N(a_k+a_i)\,{\rm log}(a_k+a_i)+ 
2a_k\,{\rm log}a_k.\label{zeroinsthea}
\ee
To evaluate the one instanton terms, we must use
\be
\frac{\pr S_k}{\pr x}(e_k) 
=S_k(a_k)\left(\sum_{i=1}^N\frac{1}{a_k+a_i}-2\sum_{i\neq k}
\frac{1}{a_k-a_i}+
\frac{2}{a_k}\right),
\ee
and eqs. (\ref{mass}) and (\ref{periods}), which imply  
\be
\sum_{i=1}^Na_i+{\cal O}(\bLa^4)=
\frac{\bLa^2}{4}\sum_{i=1}^N
\frac{\pr S_i}{\pr x}(a_i), \label{identityres}
\ee
for a massless hypermultiplet. Eq. (\ref{identityres}) vanishes as we will 
verify. To see this, introduce the function
\be
K(x)=\frac{4(-1)^Nx^2\prod_{i=1}^N(x+e_i)}{\prod_{i=1}^N(x-e_i)^2}. 
\label{helpfunc}
\ee
Notice that the residues of $K(x)$ are given by 
\be
{\rm Res}\,K(x)\mid_{x=e_k}=\frac{\pr S_k}{\pr x}(e_k),
\ee
and since $K(x)$ does not have any poles at infinity its residues must 
sum to zero. 
This gives the result 
\be
\sum_{i=1}^Na_i=0,
\ee
to this order. 

Making use of these identities, it can be checked that the $\bLa$ 
contributions 
cancel as promised, and the contribution from the hyperelliptic 
approximation to the one instanton 
term takes the form 
\be
2\pi i\,[(a_{D,k})_{1-inst}]_I\,=\, \bLa^2\left(-\frac{1}{2}\sum_{i\neq k}
\frac{S_i(a_i)}{a_k-a_i}-
\frac{1}{4}\sum_{i=1}^N\frac{\pr S_i}{\pr x}(a_i){\rm log}(a_k+a_i)+
\frac{1}{4}\frac{\pr S_k}{\pr x}(a_k)\right).
\label{oneinst}
\ee

{From} the SW ansatz we know that this result, if complete as it 
stands, must be integrable order by order in $\bLa^2$. The zero instanton 
result is exactly what one expects from the prepotential. However, 
the logarithmic term at the one-instanton level is spurious, and 
indicates that we need additional terms 
in the expansion beyond the hyperelliptic approximation. 
The corrections that we need are given by
the second term in eq.  (\ref{swdpert}).

By arguments similar to the ones used in the evaluation of the dual period 
contribution from the hyperelliptic approximation, one can convince 
oneself that the $\ep/(2 g r\bLa^2)$ 
term in (\ref{swdpert}) will not  contribute to one instanton effects. 
What remains is  $x{\rm d}(\ep r/(g^2\bLa^4))$ since only 
odd terms in $r$ 
will survive. Hence we wish to evaluate 
\be
2\pi i\,(a_{D,k})_{II}\,=\,2\bLa^2\int^{\xmk}x{\rm d}
\left(\frac{x^6r}{g^2}\right)=
2\bLa^2\int^{\xmk}{\rm d}x\left[\pr_x\left(\frac{x^7r}{g^2}\right)-
\frac{x^6r}{g^2}\right].
\ee
To this order we may take $r=f/2$. It is also easy to convince oneself 
that the 
total derivative terms are of order $\bLa^3$. This leaves us with
\be
-\bLa^2\int^{\xmk}{\rm d}x
\frac{x^2\prod_{i=1}^N(x-e_i)}{\prod_{i=1}^N(x+e_i)^2}=
-\bLa^2\sum_{i=1}^N\lef(-\frac{\bar Q_{i,2}}{x^-_k+e_i}+
\bar Q_{i,1}\,{\rm log}(x^-_k+e_i)\right),\label{alex}
\ee
where we have introduced the partial fraction coefficients 
$\bar Q_{i,p}$ for  $p=1,2$, which
 are defined by the relation
\be
\frac{x^2\prod_{i=1}^N(x-e_i)}{\prod_{i=1}^N(x+e_i)^2}\equiv
\sum_{i=1}^N\sum_{p=1}^2\frac{\bar Q_{i,p}}{(x+e_i)^p}.
\ee
Evaluating the right hand side near $x=-e_i$, one finds that
\be
\bar Q_{i,1}=-\frac{1}{4}\frac{\pr S_i}{\pr x}(e_i)\,\,\, \hs{4mm} 
{\rm and} \,\,\,\hs{4mm} 
\bar Q_{i,2}=\frac{1}{4}S_i(e_i).
\ee
The total correction term from eq. (\ref{alex}) then is 
\be
2\pi i\,(a_{D,k})_{II}\,=\,\bLa^2\sum_{i=1}^N
\left(\frac{S_i(a_i)}{4(a_k+a_i)}+
\frac{1}{4}\frac{\pr S_i}{\pr x}(a_i){\rm log}(a_k+a_i)\right).
\ee
Adding this to the one instanton contribution obtained in 
the hyperelliptic approximation, 
we at last get the complete one-instanton contribution to the 
dual period 
\be
2\pi i\,(a_{D,k})_{1-inst}\,=\bLa^2\left(-\frac{1}{2}\sum_{i\neq k}
\frac{S_i(a_i)}{a_k-a_i}+
\frac{1}{4}\sum_{i=1}^N\frac{S_i(a_i)}{a_k+a_i}+\frac{1}{4}
\frac{\pr S_k}{\pr x}(a_k)\right)-(k\rightarrow 1),
\label{finaldualp}
\ee
where we have restored the dependence on the lower integration limit which 
was suppressed throughout the calculation.

The final result for  the dual period derived from the curve (\ref{curve}) 
up to one instanton order takes  the form \np
\be
& &2\pi i a_{D,k}\,=\,2\pi i\, [(a_{D,k})_{cl}\,+\,(a_{D,k})_{1-loop}\,+
\,(a_{D,k})_{1-inst}]\nn\\
& &=[N-2+2{\rm log}(-\bLa)+{\rm log}(-1)^N]\,a_k\nn \\ 
& &-2\sum_{i\neq k}(a_k-a_i)\,
{\rm log}(a_k-a_i)+
\sum_{i=1}^N(a_k+a_i)\,{\rm log}(a_k+a_i)+ 
2a_k\,{\rm log}a_k \nn \\
& &+\bLa^2\left(-\frac{1}{2}\sum_{i\neq k}\frac{S_i(a_i)}{a_k-a_i}+
\frac{1}{4}\sum_{i=1}^N\frac{S_i(a_i)}{a_k+a_i}+\frac{1}{4}
\frac{\pr S_k}{\pr x}(a_k)\right)-(k\rightarrow 1).
\label{finalformdp}
\ee

\section{The Prepotential}

Finally we must use the dual periods to find the prepotential, i.e. to 
find the 
holomorphic function whose derivative is $a_{D,k}$. This will provide us 
with the one instanton 
correction to the prepotential for SU($N$) with matter transforming 
in the symmetric representation. 
The validity of this result can be tested 
by performing microscopic instanton calculations along the lines 
of previous calculations testing similar results for matter in the 
defining representation. 

We must show that the dual periods $a_{D,k}$ can be expressed as a 
derivative of some function with respect to 
$a_k$, as in eq. (\ref{patata}).  
We can express the prepotential as:
\be
{\cal F}\,=\,{\cal F}_{cl}\,+\,{\cal F}_{1-loop}\,+
\sum_{d=1}^{\infty}\,{\bLa}^{2d}\,{\cal F}_d. \label{express}
\ee

The one-loop part of the prepotential takes the explicit form 
\be
{\cal F}_{1-loop}=\frac{i}{8\pi}\left[\sum_{i,j=1}^N(a_i-a_j)^2\,
{\rm log}\left(\frac{a_i-a_j}{\La}\right)^2-
\!\sum_{i\leq j}(a_i+a_j)^2\,{\rm log}
\left(\frac{a_i+a_j}{\La}\right)^2\right],\,
\ee
for the symmetric representation. 
The derivative of this expression with 
respect to $a_k$ reproduces the 
functional behaviour of the 
zero instanton term of the dual periods $a_{D,k}$ c.f. 
eq. (\ref{finalformdp}). 

To check the one instanton term, first notice that 
we are interested in derivatives with respect 
to $a_k$, but in ({\ref{finalformdp}) all the derivatives 
are considered with respect to $x$. However, we have the identity 
\be
\frac{\pr S_k(a_k)}{\pr a_k}
=\frac{\pr S_k}{\pr x}(a_k)+\frac{S_k(a_k)}{2a_k}.
\ee
Furthermore 
\be
\sum_{i\neq k}\frac{S_i(a_i)}{a_k+a_i}-2\sum_{i\neq k}
\frac{S_i(a_i)}{a_k-a_i}=\frac{\pr}{\pr a_k}
\sum_{i\neq k}S_i(a_i). 
\ee
Thus, the one instanton terms of the dual periods $a_{D,k}$ can be then 
expressed as 
\be
2\pi i\,(a_{D,k})_{1-inst}&=&
\bLa^2\left[-\frac{1}{2}\sum_{i\neq k}
\frac{S_i(a_i)}{a_k-a_i}+\frac{1}{4}\sum_{i=1}^N\frac{S_i(a_i)}{a_k+a_i}+
\frac{1}{4}\frac{\pr S_k}{\pr x}(a_k)\right]\nn \\
&=&\frac{\bLa^2}{4}\frac{\pr}{\pr a_k}\sum_{i=1}^N S_i(a_i).
\ee
This demonstrates the integrability, and gives the one instanton 
correction to 
the prepotential, which is 
\be
{\cal F}_1\,=\,{1\over{8\pi i}}\,\sum_{i=1}^N S_i(a_i).
\ee
It is  straightforward 
to see that we can continue our calculation along the same lines 
to arbitrary instanton number. 
However, the number of terms will increase quickly and the calculations 
will become increasingly more difficult.

Our result for the one instanton correction ${\cal F}_1$  
must reproduce the result for pure Yang-Mills as we take the mass 
of the hypermultiplet to infinity, 
because in this limit we ``integrate out" the massive degrees of 
freedom and we are left with the 
pure gauge case only. In order to restore the mass dependence, 
we shift 
\be
a_i\lra a_i+\frac{m}{2},
\ee 
which is consistent with the mass dependence in Landsteiner and Lopez 
\cite{LandsteinerLopez}. 
In the limit $m\lra\infty$, the one 
instanton term takes the form (by using the explicit form for 
$S_i(a_i)$, c.f. (\ref{residuefunc})),
\be
\sum_{i=1}^N\frac{4(a_i+m/2)^2
\prod_{j=1}^N(a_i+a_j+m)}{\prod_{j\neq i}(a_i-a_j)^2}\lra
\sum_{i=1}^N\frac{m^{N+2}}{\prod_{j\neq i}(a_i-a_j)^2}. 
\ee 
This coincides with the result for pure Yang-Mills theory 
\cite{DHokerKricheverPhong1}, 
with appropriate 
rescalings of the respective scales $\La$. 

Notice that our expression for the one instanton correction ${\cal F}_1$ 
to the prepotential is formally identical 
to the one derived for matter transforming in the defining 
representation \cite{DHokerKricheverPhong1}, 
with the difference in the theories just expressed by 
the different definition of the residue functions $S_k$. 
This is also  
true for the antisymmetric representation 
with the minor addition of a similar term (a residue
at $x=0$) \cite{NaculichRhedinSchnitzer}.  

%Finally, the lesson to learn from previous examples that have been 
%tested by microscopic instanton calculations 
%is that any discrepancies get worse for higher instanton numbers. 
%In general, for the first instanton 
%term, one finds agreement between the mathematical SW ansatz and 
%the microscopic calculation. 

\section{Conclusions}

In this paper,  we have calculated the prepotential up to one 
instanton order for the N=2 
supersymmetric Yang-Mills theories in four dimensions with gauge group 
SU($N$) and matter transforming 
in the symmetric representation. The calculation has been performed 
with the SW 
ansatz as the starting point, together with a curve derived from M-theory. 
The curve, being cubic, is not as straightforward to deal 
with as the quadratic curves found 
for matter in the defining representation.  

It was shown that the curve reproduces the correct 
result for the one-loop part of the prepotential, which  
although a non-trivial test of the curve, is not a conclusive one. 
First of all, the curve has 
already been tested for self consistency when derived from M-theory. 
These tests might already be accurate enough to 
guarantee the one-loop result. Furthermore, as we have seen in 
this paper, the hyperelliptic approximation is 
sufficient to get the correct one-loop result, which tells us that 
there is more than one curve with 
the same one-loop prepotential. We also have unpublished results 
in which we find 
entirely different curves for SU(3) and SU(4) which also reproduce the 
correct one-loop prepotential, but fail to be integrable to one-instanton 
order.  
(This approach might be extended to SU($N$) although we have not done so.) 
One thus must go beyond one-loop to obtain non-trivial tests of 
any proposed curve. 
In fact, the one instanton term 
will distinguish between various approaches that 
reproduce the one-loop part of the prepotential. Therefore a convincing 
test of the M-theory curve requires at least the one-instanton term. 

We also gave the one instanton correction to the prepotential. 
It was found to be of the same general 
form as the one obtained from matter transforming in the defining 
representation. This is 
also the case for the antisymmetric representation, which 
is discussed in a separate 
publication \cite{NaculichRhedinSchnitzer}. 

The one instanton term is also important from the physical point of view. 
No matter how powerful 
the SW approach is, and no matter how successful it has proven in 
previous examples, it is still 
a conjecture whose results must be  tested in detailed field 
theoretic calculations. This is 
something which can be done in 
microscopic instanton calculations, and here we provide a target 
for those calculations.

\vs{2mm}

{\bf{Acknowledgement:}} We would like to thank Michael Mattis and 
\"Ozg\"ur Sar{\i}o\~{g}lu for valuable discussions. 
HJS wishes to thank the Physics Department of Harvard
University for their hospitality during the spring semester of 
1998.

$\ $\vs{5mm}

{\large\bf Appendix A}
\renewcommand{\theequation}{A.\arabic{equation}}
\setcounter{equation}{0}

%\renewcommand{\thesection}{ } 
%\renewcommand{\theequation}{A\theequation}
%\section{Appendix A}

%\setcounter{equation}{0}

Some partial results are listed below which were used to obtain 
the dual periods for the 
hyperelliptic approximation. The left-hand sides 
are taken from eq. (\ref{finalint}), and we have made extensive use of 
eq. (\ref{idthree}) and analogous expressions.  

\be
& &-2(\xmk-a_k){\rm log}(\xmk-e_k) \\
& &=2\bLa\left[{\rm log}(-\bLa)+{\rm log}\,S_k^{1/2}(a_k)\right]
S_k^{1/2}(a_k)-
\bLa^2\left[1+\frac{1}{2}{\rm log}(-\bLa)+
\frac{1}{2}{\rm log}\,S_k^{1/2}(a_k)\right]\frac{\pr S_k}{\pr x}(a_k).\nn 
\ee

\be
& &-2\sum_{i\neq k}(\xmk-a_i){\rm log}(\xmk-e_i)\\
& &=-2\sum_{i\neq k}(a_k-a_i){\rm log}(a_k-a_i)+2\bLa\left[N-1+
\sum_{i\neq k}{\rm log}(a_k-a_i)\right]
S_k^{1/2}(a_k) \nn \\
& &-\bLa^2\left[\frac{N-1}{2}\frac{\pr S_k}{\pr x}(a_k)+
\frac{1}{2}\frac{\pr S_k}{\pr x}(a_k)\sum_{i\neq k}{\rm log}(a_k-a_i)+
\frac{1}{2}\sum_{i\neq k}\frac{\pr S_i}{\pr x}(a_i)+
S_k(a_k)\sum_{i\neq k}\frac{1}{a_k-a_i}\right].\nn 
\ee

\be
& &\sum_{i=1}^N(\xmk+e_i){\rm log}(\xmk+e_i) \\
& &=\sum_{i=1}^N(a_k+a_i){\rm log}(a_k+a_i)-\bLa\left[N+
\sum_{i=1}^N{\rm log}(a_k+a_i)\right]S_k^{1/2}(a_k) \nn \\
& &+\bLa^2\left[
\frac{N}{4}\frac{\pr S_k}{\pr x}(a_k)
+\frac{1}{4}\frac{\pr S_k}{\pr x}(a_k)\sum_{i=1}^N{\rm log}(a_k+a_i)-
\frac{1}{4}\sum_{i=1}^N\frac{\pr S_i}{\pr x}(a_i){\rm log}(a_k+a_i) 
\right. \nn \\
& &-\left.\frac{1}{4}\sum_{i=1}^N\frac{\pr S_i}{\pr x}(a_i)+
\frac{1}{2}S_k(a_k)\sum_{i=1}^N\frac{1}{a_k+a_i}\right]. \nn
\ee

\be
& &2\xmk{\rm log}\xmk\\
& &=2a_k{\rm log}a_k-2\bLa S_k^{1/2}(a_k)(1+{\rm log}\,a_k)+
\bLa^2\left[\frac{1}{2}\frac{\pr S_k}{\pr x}(a_k)(1+{\rm log}\,a_k)
+\frac{S_k(a_k)}{a_k}\right]. \nn 
\ee

\end{document}